\documentclass[twocolumn,prb]{revtex4}
\usepackage{amsmath}
\usepackage{amsfonts}
\usepackage{amssymb}
\usepackage{graphics}
\usepackage{float}
\usepackage{graphicx}
\usepackage{epstopdf}
\usepackage[compact]{titlesec}
\usepackage{natbib}
\usepackage{threeparttable}
\usepackage[titletoc,title]{appendix}
\usepackage{lipsum}
\usepackage{textcomp}
\usepackage{subfigure}

\begin{document}
\title{The optical response of mono-layer transition metal dichalcogenides embedded in a Kerr-type non-linear dielectric environment}
\author{Parijat Sengupta$^{1}$}
\email{parijats@bu.edu}
\author{Shaloo Rakheja$^{2}$}
\author{Enrico Bellotti$^{1}$}
\affiliation{$^{1}$ Photonics Center, Boston University, Boston, MA, 02215 \\
$^{2}$ Electrical and Computer Engineering, New York University, New York, NY, 11201 
}

\begin{abstract}
We study the optical behaviour of an arrangement in which the interface between a linear and non-linear dielectric media is covered by an embedded mono-layer of transition metal dichalcogenides (TMDC). The optical behaviour is qualitatively obtained through transmission and reflection coefficients which are a function of the third order non-linear susceptibility of the Kerr-type dielectric and the inter-band optical conductivity of the TMDC mono-layer. The inter-band optical conductivity of the TMCD mono-layer is calculated using the Kubo formalism from linear response theory. In particular, we theoretically demonstrate that the optical response of this arrangement can be switched between total internal reflection and a normal transmission regime by controlling the intensity of the incident radiation. The reflection and transmission functions, additionally, are shown to be amenable to further control by altering the inter-band optical conductivity of the embedded TMDC mono-layer. The optical conductivity is an outcome of its energy dispersion; we specifically choose two TMDC mono-layers, MoS$_{2}$ and WSe$_{2}$, which have nearly identical dispersion parameters apart from a much stronger spin-orbit coupling in the latter. The stronger spin-orbit coupling in WSe$_{2}$ does not significantly alter the inter-band optical conductivity to manifest in an enhanced reflection spectrum. However, we find that application of an external perturbation such as strain could be effectively used to modulate the overall optical response. We conclude by offering a brief overview of the applicability of our proposed scheme in devices that employ an all-optical switching mechanism through optical bistability which is the hallmark of a non-linear dielectric. 
\end{abstract}
\maketitle
\vspace{0.35cm}
\section{Introduction}
\vspace{0.35cm}
The transition metal dichalcogenides (TMDCs) are layered materials of covalently bonded atoms held together by weak van der Waals forces~\cite{wilson1969transition,wang2012electronics}. A representative compound of the TMDC family, molybdenum disulphide (MoS$_{2}$), for example, consists of layers of molybdenum atoms surrounded by sulphur in a trigonal prismatic arrangement~\cite{ramakrishna2010mos2}. The bulk MoS$_{2}$ is an indirect band gap semiconductor which can, however, through mechanical exfoliation yield a layered two-dimensional configuration of Mo and S atoms. This two-dimensional arrangement or a single layer of MoS$_{2}$ is a direct band gap semiconductor $\left(E_{g} = 1.85\,eV\right) $ unlike its bulk counterpart.~\cite{chhowalla2013chemistry,huang2013metal} 
Further, in contrast to the semi-metallic, two-dimensional (2D) graphene~\cite{novoselov2005two}, the direct band gap of MoS$_2$ makes it optically active in the visible electromagnetic spectrum~\cite{splendiani2010emerging} positioning it as a potential complement to graphene-based photonic applications. In addition, the mono-layer variants of TMDCs exhibit a great degree of sensitivity to external influences, for example, strain which significantly alters their band gap and the overall electron spectrum allowing for variable light absorption suitable for a wide class of applications. While idealized optical set-ups with TMDs have been investigated before~\cite{britnell2013strong,peng2015two,eda2013two,jiang2015broadband}, the ability to control the electromagnetic response of the media through which light wave travels and the attenuation it suffers before being absorbed for purposes of detection and modulation can yield useful results and aid the design of better optical devices. 

In this work, we account for these phenomena and obtain absorption characteristics of a mono-layer TMDC sheet placed between two dielectrics as a function of parameters that define the electromagnetic response of a medium. Note that while MoS$_{2}$ is selected as our representative TMDC, we also carry out numerical calculations with other commonly known compounds of this family such as WSe$_{2}$ and compare their optical behaviour. To begin, we consider a model structure wherein the MoS$_{2}$ sheet is located at the interface of two dielectric films and show using the transfer matrix formalism that the optical response of embedded MoS$_{2}$ is largely dependent on the nature of dielectrics, incident angle, and wavelength of the incident electromagnetic (EM) wave. This dependence is quantitatively expressed through the transmission and reflection spectrum using transfer matrices that relate the amplitudes of electric and magnetic fields of the incident EM wave propagating across the two optical media separated by the MoS$_{2}$ sheet. We stress that the nature of dielectrics come in to play since their intrinsic non-linearity is explicitly accounted for in our calculations. It is worthwhile to note that while every dielectric demonstrates a certain degree of non-linearity in their response to an electromagnetic disturbance, for most part the corrections are ignored; however, to explain phenomena, for instance, optical bistability~\cite{gibbs2012optical,lugiato1984theory} and hysteresis of reflected light~\cite{kaplan1977theory,afanas1998hysteresis}, it becomes necessary to seek recourse to an EM field dependent dielectric constant.~\cite{mihalache1987exact,chen1993exact}

The rationale behind a survey of the non-linear optical properties is grounded in the fact that it paves the way for tuning the behaviour of 2D materials to modulate the amplitude of an optical signal for desired photonic applications. While a diverse set of photonic functionalities such as reverse saturable absorption, two-photon and free-carrier absorption, and nonlinear refraction can be achieved through a dynamic control employing an active feedback mechanism, the inherent non-linearity as part of the physical characteristic of the material ensures that response times to an EM disturbance are not hindered by a low rate of signal transmission between the foundational modules. The overall device architecture with passive design components can potentially be very simple and fast and crucial for manipulating short optical pulses. For a detailed overview of the practicality of optical structures whose operation is aided by non-linear behaviour, we refer the reader to Ref.~\onlinecite{tutt1993review}.

To keep things tractable and amenable to analytic derivations, the dielectrics in our model are assumed to conform to the Kerr-type classification.~\cite{gong2011highly,bottcher1978theory}. We also note that in addition to the response of the optical medium, the electronic structure of the TMDC mono-layer marks its imprint on the passage of an incident electromagnetic wave by virtue of its intrinsic surface conductivity; an estimate of the conductivity of TMDC mono-layer for use in our calculations is performed in the inter-band regime using the Kubo formalism from linear response theory.~\cite{mahan2013many} It is relevant to mention however, that several low-energy optical transitions in mono-layer TMDCs are attributed to excitons (mostly bound to impurities) which are relatively stable in these materials because of high hole and electron effective masses. However, in this paper we focus exclusively on optical conductivity that arises from an inter-band transition in a pristine TMDCs mono-layer and choose to ignore the excitonic contributions. 

Throughout this study, to evaluate the optical behaviour of our proposed arrangement, we draw attention of the reader to 1) the role of non-linear coefficients that characterize the dielectrics, 2) the electronic structure of the mono-layer TMDC sheet~\cite{li2015symmetry,kormanyos2013monolayer} that absorbs a part of the incident wave underscoring its value in regulating the inter-band optical conductivity and 3) possible design metrics that could enhance light-matter interaction~\cite{lundeberg2014harnessing} in transition metal dichalcogenide nano-sheets. In particular, we theoretically demonstrate that there exists a critical angle which is intensity-dependent and is conspicuous by a sharp dip in the reflection coefficient. The sharp dip marks the onset of normal transmission from total internal reflection as the intensity of the incident light beam goes beyond the threshold value. The passage of light is also shown to be governed by the inter-band optical conductivity (which is a direct outcome of the energy dispersion) of the embedded TMDC mono-layer, hence, for a comparative examination, we choose for the embedded layer MoS$_{2}$ and WSe$_{2}$ as representative TMDCs. These compounds which have almost identical band parameters but a widely different spin-orbit coupling does not manifest as enhanced inter-band optical conductivity and the overall percentage of reflected and transmitted light stays practically unaltered. However, seeking recourse to an external perturbation such as bi-axial mechanical strain in the TMDC mono-layer offers additional control by modulating the reflected and transmitted light and also adjusts the intensity-dependent critical angle. 

The attenuation and transmission of light is therefore not only an upshot of ambient conditions and other vital parameters including the angle of incidence, but also, crucially depends on a strong light-matter interaction. The TMDC nanostructure. A closing summary offers suggested enhancements to the presented work, especially in connection to the Kerr non-linearity assisted phenomenon of optical bistability that is actively pursued as a pathway to greater integration of nano-photonic circuits with mainstream semiconductor electronics. Proofs of analytic calculations left out in the main body of the text are collected in two appendices.

\vspace{0.35cm}
\section{Model and Theory}
\begin{figure}
\centering
\includegraphics[width=3.3in]{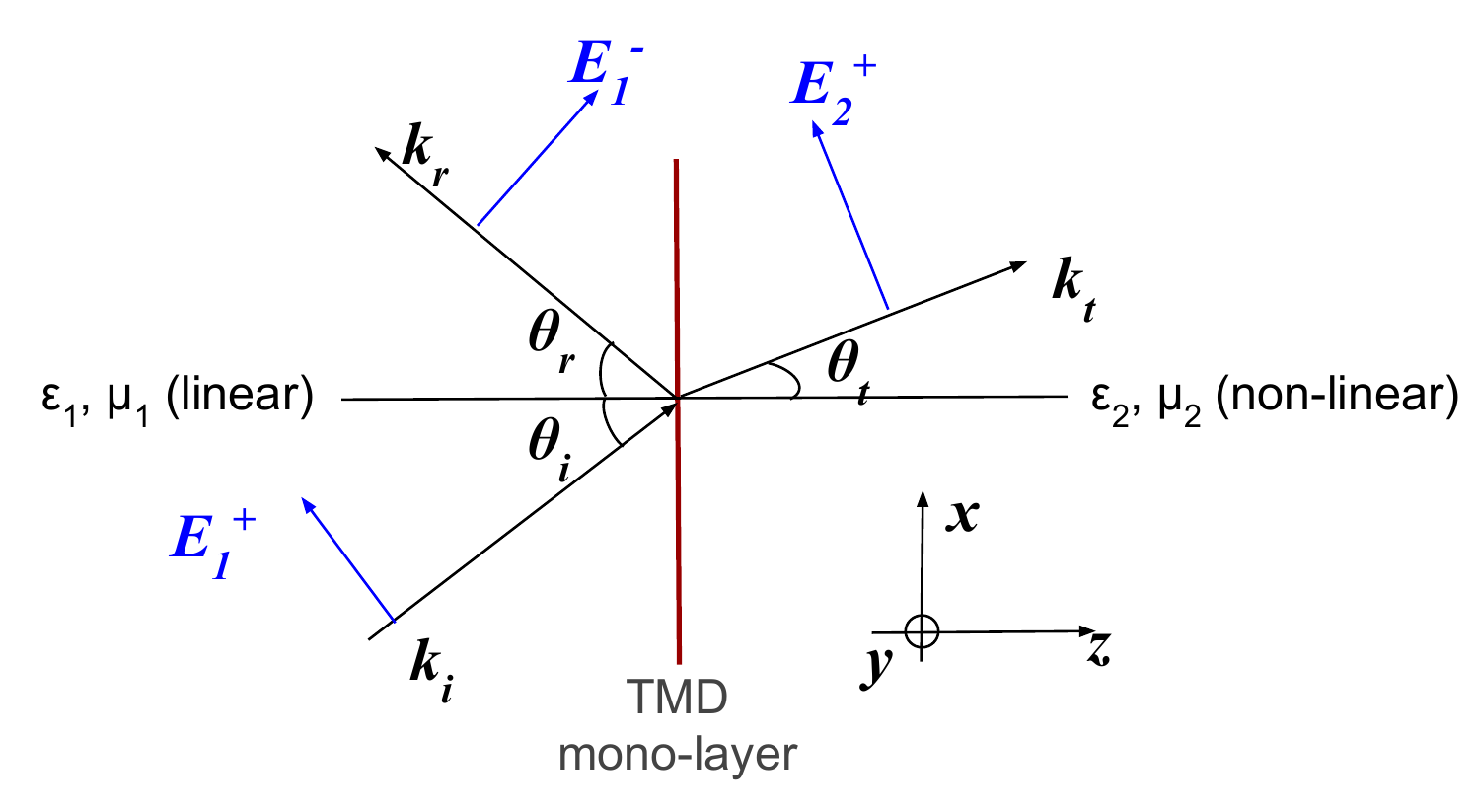} 
\caption{Schematic of the MoS$_{2}$ mono-layer inserted at the interface of a linear and non-linear dielectric. The linear (non-linear) dielectric is identified as $ \epsilon_{1}\left(\epsilon_{2}\right)$. The non-linearity of the dielectric is assumed to be of the Kerr-type.}
\label{schema}
\end{figure}

\vspace{0.35cm}
We first sketch the structure in Fig.~\ref{schema} of an embedded MoS$_{2}$ mono-layer as the representative TMDC in a dielectric environment. A light wave of certain frequency $ \omega $ is incident on the structure from left and exits after partial absorption in the MoS$_{2}$ mono-layer. The dielectric on the left is linear while the MoS$_{2}$ sheet to its right is flanked by a Kerr (non-linear) dielectric. Our theory calculations are divided in two sub-sections; we first work out the energy spectrum of the TMDC mono-layer around the $ K $ and the $ K^{'} $ valley edge of the Brillouin zone and use it to determine the inter-band optical conductivity followed by a transfer matrix formalism approach to quantitatively establish analytic expressions for the reflectance and transmittance coefficient when a light beam  is incident on the TMDC sheet separating two optical media (see Fig.~\ref{schema}). The inter-band optical conductivity is required for imposition of proper electromagnetic boundary condition as we show later (Section.~\ref{kerrd}). We explicitly declare that all optical phenomena happen in the vicinity of the $ K $ and the $ K^{'} $ valley edge, which are high-symmetry points.   

\vspace{0.35cm}
\subsection{Optical conductivity of mono-layer transition metal dichalcogenides}
\vspace{0.35cm}
The basis for the optical conductivity calculation for TMDCs such as MoS$_{2}$ (which, as before, serves as the typical TMDC) is the low-energy Hamiltonian~\cite{xiao2012coupled}
\begin{equation}
H_{\tau} = a\,t\left(\tau k_{x}\hat{\sigma}_{x} + k_{y}\hat{\sigma}_{y}\right)\otimes \mathbb{I} + \dfrac{\Delta}{2}\hat{\sigma}_{z}\otimes \mathbb{I} - \dfrac{\lambda\,\tau}{2}\left(\hat{\sigma}_{z}-1\right)\otimes\,\hat{s}_{z}. 
\label{mos2ham}   
\end{equation}
The index $ \tau $ distinguishes the two valley edges $ K $ and $ K^{'} $, $ a $ is the lattice constant, and $ t $ denotes the hopping parameter. The energy gap between the conduction and valence bands is $ \Delta $ and the spin-orbit coupling is $ 2\lambda $. The Pauli matrices $ \hat{\sigma}_{i} $ where $ i = \left\lbrace x, y, z\right\rbrace $ act on the basis sub-space while $ \hat{s}_{z} $ is linked to the spin of the electrons at the $ K $ and $ K^{'} $ points. Quadratic terms in the Hamiltonian and trigonal warping are not included in the conductivity calculations since corrections arising out of them can be neglected for small values of $ \textbf{k} $. Note that the non-equivalent high-symmetry points $ K $ and $ K^{'} $  are degenerate and related through time reversal symmetry.~\cite{zhu2014study,konabe2014valley} Also, the Hamiltonian in Eq.~\ref{mos2ham} for mono-layer TMDC can be split in to a conduction and valence part by expanding the matrices as follows 
\begin{subequations}
\begin{equation}
H_{\tau} = \begin{pmatrix}
\Delta/2 & at\tau\,ke^{-i\theta} & 0 & 0 \\
at\tau\,ke^{i\theta} & -\Delta/2  + \lambda\,\tau & 0 & 0 \\
0 & 0 & \Delta/2 & at\tau\,ke^{-i\theta} \\
0 & 0 & at\tau\,ke^{i\theta} & -\Delta/2  - \lambda\,\tau
\end{pmatrix},
\label{expham}
\end{equation}
where $ \theta = tan^{-1}\dfrac{k_{y}}{k_{x}} $. The subscript $ \tau = \pm $ denotes the $ K $ and $ K^{'} $ valley edge of the mono-layer TMDC. Further, the upper (lower) 2 $\times $ 2 block in Eq.~\ref{expham} is the set of spin-up (down) conduction and valence bands separated by a band-gap of $ \Delta $. The eigen energies and wave functions are easily obtained for the Hamiltonian in Eq.~\ref{expham} as shown below. For all calculations from now on, we only use the $ K $ edge and therefore drop the subscript $ \tau $ and set it to unity everywhere.
\begin{equation}
\begin{aligned}
E_{\uparrow} &= \dfrac{1}{2}\biggl[\lambda \pm \sqrt{\left(\Delta - \lambda\right)^{2} + 4a^{2}t^{2}k^{2}}\biggr], \\
E_{\downarrow} &= \dfrac{1}{2}\biggl[-\lambda \pm \sqrt{\left(\Delta + \lambda\right)^{2} + 4a^{2}t^{2}k^{2}}\biggr].
\label{eigval}
\end{aligned}
\end{equation}
The $ +\left(-\right) $ in the eigen energy expressions correspond to the conduction (valence) band. Note that the finite spin-orbit coupling ($ 2\lambda $) splits the valence bands at $ K $ while the conduction states remain spin degenerate. The eigen values in vicinity of the $ K $ valley-edge for mono-layer MoS$_{2}$ are plotted in Fig.~\ref{eigvalpic}. In obtaining this dispersion plot, we have used the material constants collected in Table.~\ref{table1} for Mo$S_{2}$. The parameters of WSe$_{2}$ are also listed for later use. 
\begin{table}
\caption{Band structure parameters for representative TMDCs MoS$_{2}$ and WSe$_{2}$.~\cite{xiao2012coupled}}
\centering
\label{table1}
\begin{tabular}{lcc}
\noalign{\smallskip} \hline \hline \noalign{\smallskip}
Parameters & MoS$_{2}$  & WSe$_{2}$\\\hline
a(\AA) & 3.193 & 3.310 \\
$ \Delta\,(eV) $ & 1.66 & 1.60 \\
$ t\,(eV) $ & 1.10 & 1.19 \\
$ 2\lambda\,(eV) $ & 0.15 & 0.46 \\
\noalign{\smallskip} \hline \noalign{\smallskip}
\end{tabular}
\end{table}
\begin{figure}
\centering
\includegraphics[scale=1]{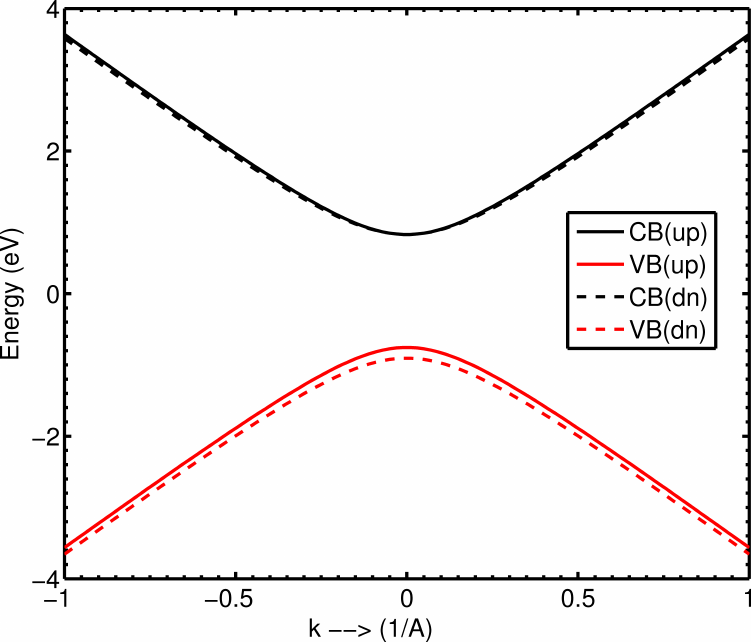} 
\caption{The band dispersion of mono-layer Mo$S_{2}$ at the $ K $ point $(\tau = 1)$. The bands are spin-resolved. Note that spin-orbit coupling pushes the spin-down component of the valence band and they are no longer degenerate at $ K $ contrary to the spin-states of the conduction band. The effective band gap (energy gap between the spin-up conduction and valence states) at $ K $ on account of spin-orbit coupling is now $ \Delta - \lambda $.}
\label{eigvalpic}
\end{figure} 
The band gaps between the spin-up conduction and valence bands and their counterpart spin-down states using Eq.~\ref{eigval} are
\begin{equation}
\begin{aligned}
\Delta\,\varepsilon_{up} &= \sqrt{\left(\Delta - \lambda\right)^{2} + 4a^{2}t^{2}k^{2}},  \\
\Delta\,\varepsilon_{down} &= \sqrt{\left(\Delta + \lambda\right)^{2} + 4a^{2}t^{2}k^{2}}.
\label{bgs}
\end{aligned}
\end{equation}

\noindent We also write down the eigen spinors for the conduction and valence bands for mono-layer TMDC which take the form
\begin{equation}
\Psi^{up}_{\pm} = \dfrac{1}{\sqrt{2}}\begin{pmatrix}
\eta_{\pm}e^{-i\theta} \\
\pm\,\eta_{\mp}
\end{pmatrix},
\label{wfs1}
\end{equation}
where 
\begin{equation}
\eta_{\pm}^{up} = \sqrt{1 \pm \dfrac{\Delta - \lambda}{\left(\Delta - \lambda\right)^{2} + \left(2atk\right)^{2}}} .
\label{etaexp1}
\end{equation}
The subscript $ + (-) $ stands for the conduction (valence) state. These wave functions are derived by diagonalizing the upper $ 2 \times 2 $ block of the Hamiltonian in Eq.~\ref{expham} which represents the spin-up component for the conduction and valence bands. Note that we can derive a similar set of wave functions for the spin-down components by choosing the lower $ 2 \times 2 $ block. We only need to replace the $ \Delta - \lambda $ in Eq.~\ref{wfs1} with $ \Delta + \lambda $ to yield eigen spinors $ \Psi^{down}_{\pm} $ of the form
\begin{equation}
\eta_{\pm}^{down} = \sqrt{1 \pm \dfrac{\Delta + \lambda}{\left(\Delta + \lambda\right)^{2} + \left(2atk\right)^{2}}} .
\label{etaexp2}
\end{equation}
\end{subequations}
In passing we also write down the velocity operator based on the Hamiltonian in Eq.~\ref{mos2ham}; the velocity operator is $ \hat{v}_{x} = \dfrac{1}{i\hbar}\left[\hat{x}, \mathcal{H}_{\tau}\right] $ which gives $ \hat{v}_{x} = \dfrac{at}{\hbar}\hat{\sigma}_{x}$.
We have now acquired all the desired pieces to compute the optical conductivity. In what follows, we ignore the intra-band contribution and focus exclusively on the inter-band component to obtain the frequency-dependent longitudinal conductivity $ \sigma_{xx} $ of a mono-layer TMDC.

\begin{figure}
\centering
\includegraphics[width=3.4in]{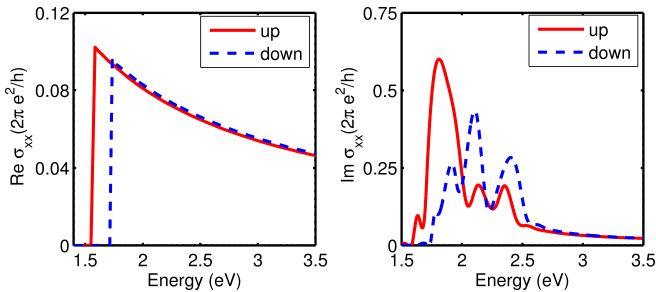} 
\caption{The spin-resolved longitudinal real (left panel) and imaginary part of the optical conductivity of mono-layer MoS$_{2}$ at $ T = 0\,K $ in vicinity of the $ K $ point of the Brillouin zone. The sheet is assumed to be hole-doped such that the Fermi energy is aligned to top of the spin-up valence band at the $ K $ point. The conductivity numbers are obtained based on the optical transition gap that exists between the spin-up and spin-down components of the conduction and valence bands. The momentum cutoff was set at $ 0.5 \,1/\AA $ around the $ K $ point.}
\label{inbc}
\end{figure}

The optical conductivity of a non-interacting sample is expressed by the Kubo formula from linear response theory~\cite{jishi2013feynman,mahan2013many} 
\begin{equation}
\sigma_{x,y} = -i\dfrac{\hbar\,e^{2}}{L^{2}}\sum_{n,n^{'}}\dfrac{f\left(\varepsilon_{n}\right)- f\left(\varepsilon_{n^{'}}\right)}{\varepsilon_{n} - \varepsilon_{n^{'}}}\dfrac{\langle\,n\vert\,v_{\alpha}\vert\,n^{'}\rangle \langle\,n^{'}\vert\,v_{\alpha}\vert\,n\rangle}{\varepsilon_{n} - \varepsilon_{n^{'}}+i\,\zeta},
\label{kubof}
\end{equation}
where $ \vert\,n\rangle $ and $ \vert\,n^{'}\rangle $ are eigen functions of the Hamiltonian given in Eq.~\ref{mos2ham} and $ \zeta $ represents a finite broadening of the {eigen-states} resulting from surface imperfections. We note that the wave functions are still assumed to retain the form shown in Eq.~\ref{wfs1}. Inserting the wave functions from Eq.~\ref{wfs1} and the appropriate velocity components in Eq.~\ref{kubof}, we determine an expression for inter-band conductivity. The MoS$_{2}$ sample area is assumed to be $ \mathcal{A} = L^{2} $. 
\begin{subequations}
Expanding the Kubo formula for inter-band conductivity, we get (the detailed calculations are shown in Appendix B) 
\begin{align}
\begin{split}
\sigma_{xx}&= -i\dfrac{e^{2}}{\hbar}\dfrac{1}{L^{2}}\sum_{n,n^{'}}\dfrac{a^{2}t^{2}\left(\Upsilon^{2}cos^{2}\theta + sin^{2}\theta\right)}{\Delta\,\varepsilon} \\
&\times \biggl[\dfrac{1}{\hbar\,\omega + \Delta\,\varepsilon + i\zeta} 
+ \dfrac{1}{\hbar\,\omega - \Delta\,\varepsilon + i\zeta}\biggr],
\end{split}
\end{align}
where 
\begin{equation}
\Delta\,\varepsilon = \sqrt{\left(\Delta \pm \lambda\right)^{2} + 4a^{2}t^{2}k^{2}},
\label{dele}
\end{equation}
and
\begin{equation}
\Upsilon = {\dfrac{\Delta \pm \lambda}{\sqrt{\left(\Delta \pm \lambda\right)^{2} + \left(2atk\right)^{2}}}}.
\label{matelm} 
\end{equation}
We assume the Fermi distribution to be a Heaviside function such that $ f\left(\varepsilon_{c}\right) = 0 $ and $ f\left(\varepsilon_{v}\right) = 1 $. Also, notice that the matrix element $ \mathcal{M} = \langle\,n\vert\,\hat{v}_{x}\vert\,n^{'}\rangle $ evaluates to $ \mathcal{M} = -at\left(Acos\theta + isin\theta\right) $ and we are working with optical band gaps $\left(\Delta\,\varepsilon = \Delta\,\varepsilon_{up/down}\right)$ between the spin-up and spin-down states. Changing the summation over to an integral and using the standard identity~\cite{hassani2013mathematical} $ \dfrac{1}{x + i0^{+}} = \mathbb{P}\dfrac{1}{x} - i\pi\delta\left(x\right)$ allows us to write the real and imaginary terms for the longitudinal conductivity 
\begin{align}
\begin{split}
\sigma_{xx}^{\mathfrak{R}} &= \dfrac{e^{2}}{4\pi\hbar}\int_{0}^{k_{c}}k\,dk\int_{0}^{2\pi}d\theta\,\dfrac{a^{2}t^{2}\left(\Upsilon^{2}cos^{2}\theta + sin^{2}\theta\right)}{\Delta \,\varepsilon} \\
&\times \left\lbrace \delta\left(\hbar\omega + \Delta\varepsilon\right) +  \delta\left(\hbar\omega - \Delta\varepsilon\right)\right\rbrace,  
\label{rsigxx}
\end{split}
\end{align}
and
\begin{align}
\begin{split}
\sigma_{xx}^{\mathfrak{Im}} &= \dfrac{e^{2}}{4\pi^{2}\hbar}\int_{0}^{k_{c}}k\,dk\int_{0}^{2\pi}d\theta\,\dfrac{a^{2}t^{2}\left(\Upsilon^{2}cos^{2}\theta + sin^{2}\theta\right)}{\Delta \,\varepsilon} \\
&\times \left\lbrace \mathbb{P}\dfrac{2\hbar\omega}{\left(\hbar\omega\right)^{2} - \left(\Delta\varepsilon\right)^{2}}\right\rbrace.
\label{imsigxx}
\end{split}
\end{align}

A numerical estimate of the real and imaginary components of the longitudinal conductivity is needed for calculating the reflectance and transmittance coefficients derived later in Eqs.~\ref{rcff} and ~\ref{tcff}) of the following section. To evaluate the real component of the conductivity using Eq.~\ref{rsigxx}, we observe that the first $ \delta(\cdot) $ function vanishes everywhere(for a non-vanishing contribution, the argument of $ \delta\left(\hbar\omega + \Delta\varepsilon\right) $ must be zero which implies a negative frequency and is physically implausible). The upper-limit, $ k_{c} $, on the integrals denotes the momentum cutoff which is set to $ 0.5\, 1/\AA $. Note that since we employ a k.p Hamiltonian valid in the vicinity of the $ K $ and $ K^{'} $ point, high-energy regions are inaccessible in this model. The final spin-resolved expression for the real part of the longitudinal conductivity (after substituting for $ \Upsilon $ from Eq.~\ref{matelm}) is therefore 
\begin{equation}
\sigma_{xx}^{\mathfrak{R}} = \dfrac{a^{2}t^{2}e^{2}}{4\hbar^{2}\omega}\int_{0}^{k_{c}}\left(1 + \dfrac{\left(\Delta \pm \lambda\right)^{2}}{\left(\Delta \pm \lambda\right)^{2} + \left(2atk\right)^{2}} \right)k\,dk.
\label{rsigxx1}
\end{equation}
The angular part in Eq.~\ref{rsigxx} is integrated out using $ \int_{0}^{2\pi}sin^{2}\theta \, d\theta = \pi $. 
\end{subequations}
The lower (upper) sign gives the inter-band conductivity for the spin-up (down) states. The integral in Eq.~\ref{rsigxx1} is evaluated numerically to obtain the real part of the inter-band conductivity. The imaginary part of the conductivity is calculated likewise using Eq.~\ref{imsigxx} for optical transitions between spin-up and spin-down states. Notice that the imaginary part of the optical conductivity (right panel of Fig.~\ref{inbc}) shows a sharp rise at energies that correspond to a possible inter-band optical transition. The conductivity components can be seen as spin-resolved since they yield two set of numbers, each number corresponding to a fundamental optical transition gap from one of the spin-split valence bands to the vicinity of the bottom of the conduction band. 

\begin{figure}
\centering
\includegraphics[width=2.6in]{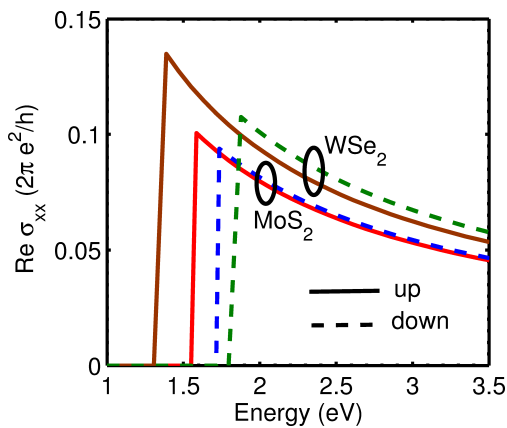}
\caption{The comparative real part of optical longitudinal conductivity for MoS$_{2}$ and WSe$_{2}$. The optical transition gap between spin-up conduction and valence band in WSe$_{2}$ is lower by approximately 0.21 $\mathrm{eV} $ than MoS$_{2}$ which translates in to a higher optical conductivity for the former. All calculations are done with the same set of numerical values noted in Fig.~\ref{inbc}.}
\label{realcomplc}
\end{figure}

The role of spin-orbit splitting in determination of the longitudinal conductivity components can be easily seen by overlaying the numbers for WSe$_{2}$ on that of MoS$_{2}$. MoS$_{2}$ and WSe$_{2}$ as semiconducting TMDCs have nearly identical band parameters (see Table.~\ref{table1}) but with a three-fold reduced spin splitting for the former. The comparative figure is shown in Fig.~\ref{realcomplc} where the lower optical transition gap (by roughly 0.21 $\mathrm{eV} $) in WSe$_{2}$ between the spin-up conduction and valence bands at $ K $ furnishes a higher value for the real part of the optical longitudinal conductivity.

We next utilize the optical conductivity obtained in this sub-section in fixing the boundary conditions that establishes the reflectance and transmittance coefficients for an incident EM wave propagating through the layered media configuration of Fig.~\ref{schema}. Notice that an external strain~\cite{guo2014tuning,pearce2015tight} rearranges the lattice and results in a different value of the hopping parameter $ ``t" $ and the $ K $-edge band gap, $ \Delta $ in Eq.~\ref{mos2ham}, to alter the inter-band optical conductivity.

\vspace{0.35cm}
\subsection{Kerr-type non-linear dielectric}
\vspace{0.35cm}
\label{kerrd}
The dielectric to the right of the TMDC mono-layer (Fig.~\ref{schema}) which is weakly non-linear and of the Kerr-type is characterized by the following relation~\cite{fibich2015the}
\begin{equation}
\\n_{nl} = n_{0} + n_{2}\vert E\left(r,t\right)\vert^2,
\label{knl}
\end{equation}   
where $ n_{nl} $ is the non-linear refractive index and $ n_{2} $ is the Kerr coefficient. In this representation, note that the electric field acts on itself through the non-linearity.~\cite{boyd2003nonlinear} For our case, we consider an incident TM wave whose electric and magnetic fields are given by Eq.~\ref{emfields} and derive expressions for the reflectance and transmittance coefficients. 
\begin{flalign}
{H}_{y,i} &= \left(A_{i}e^{-ik_{z,i}z} + B_{i}e^{ik_{z,i}z}\right)e^{ik_{x}x}, \notag \\
E_{z,i} &= \frac{k_x}{\omega\epsilon_0\epsilon_i}\left(A_ie^{-ik_{z,i}z}+B_ie^{ik_{z,i}z}\right)e^{ik_xx}, \notag \\
E_{x,i} &= \frac{k_{z,i}}{\omega\epsilon_0\epsilon_i}\left(A_ie^{-ik_{z,i}z}-B_ie^{ik_{z,i}z}\right)e^{ik_xx},
\label{emfields}
\end{flalign}
where $ k_{z} = k_{in}\cos\theta $ and $ k_{x} = k_{in}\sin\theta $. A particular physical quantity within the \textit{i$^{th}$}layer in the structure is identified by the subscript \textit{i}, $ k_{in} $ is incident wave vector, and the angle of incidence is $ \theta $. Note that the magnetic field of the incident light of frequency $ \omega $ is directed along the \textit{y}-axis.
However, the analytic forms of the electric and magnetic fields are no longer plane wave solutions in the non-linear dielectric and must be determined by solving a Helmholtz wave equation of the form (see Appendix A for a complete derivation)
\begin{subequations}
\begin{equation}
\dfrac{\partial^{2}H_{y}}{\partial\,z^{2}} + k_{z}^{2}H_{y} = 0.
\label{nl1}
\end{equation}
Rewriting Eq.~\ref{nl1} with $ k_{z}\,= \sqrt{\left(\omega^{2}/c^{2}\right)\epsilon_{nl} - k_{x}^{2}} $ and inserting the non-linear polarization~\cite{ramakrishna2008physics, boyd2003nonlinear} as $ \chi^{\left(3\right)}_{m}\vert H_{2,y}\vert^{2}H_{2,y} $, the Helmholtz equation in non-linear medium (identified as medium 2 and all pertinent variables in this medium are identified with a subscript ``2'') can be written as
\begin{equation}
\dfrac{\partial^{2}H_{2,y}}{\partial\,z^{2}} + \dfrac{\omega^{2}}{c^{2}}\left(\epsilon_{2,l} + \chi^{\left(3\right)}_{m}\epsilon_{2,l}\vert H_{2,y}\vert^{2} - \epsilon_{1}sin^{2}\theta \right)H_{2,y}= 0, 
\label{nl2}
\end{equation}
where the simple dielectric constant of medium ``2" without the non-linear part is $ \epsilon_{2,l} $. In writing the above form for the non-linear Helmholtz equation (Eq.~\ref{nl2}), we follow Ref.~\onlinecite{ramakrishna2008physics, boyd2003nonlinear} to account for the non-linear polarization of the medium that acts on itself through the term $ \chi^{\left(3\right)}_{m}\vert H_{2,y}\vert^{2} $. The constant $ \chi^{\left(3\right)}_{m} $ is the third-order magnetization non-linearity coefficient.~\cite{selfnote} Notice that this non-linear polarization is similar in form to that contained in Eq.~\ref{knl}. Also, $ k_{x}^{2} = \dfrac{\omega^{2}}{c^{2}}\epsilon_{1}sin^{2}\theta $ is the component along the interface that remains unchanged and $ \epsilon_{2}^{0} $ is the dielectric constant of the non-linear medium ``2".
\end{subequations}
Following Ref.~\onlinecite{ramakrishna2008physics}, Eq.~\ref{nl2} can be solved by assuming an ansatz of the form $ H\left(z\right) = A\,sech\left[\kappa_{z}\left(z -z^{'}\right)\right] $ where $ z^{'} $ is a variable determined from boundary conditions. Inserting the ansatz in Eq.~\ref{nl2}, gives    
\begin{subequations}
\begin{equation}
A = \kappa_{z}\sqrt{\dfrac{2}{\chi^{\left(3\right)}_{m}\epsilon_{2,l}}}, 
\label{nl3}
\end{equation}
and
\begin{equation}
\kappa_{z}^{2} = k_{x}^{2} - \epsilon_{2}^{0}\dfrac{\omega^{2}}{c^{2}}.
\label{nl4}
\end{equation}
The variable $ z^{'} $ which is a constant of integration is evaluated by noting the magnetic field amplitude at the interface $\vert H_{y,int}\vert $. The solution to the magnetic field equation in non-linear Kerr media is therefore
\begin{equation}
H_{2,y} = \sqrt{\dfrac{2}{\chi^{\left(3\right)}_{m}\epsilon_{2,l}}}\kappa_{z}\,sech\left[\kappa_{z}\left(z - z^{'}\right)\right].  
\label{nl5}
\end{equation}
Using the continuity equations for the tangential components at the interface $\left(z = 0\right) $, the constant evaluates to
\begin{equation}
\kappa_{z}z^{'} = cosh^{-1}\left(\sqrt{\dfrac{2}{\chi^{\left(3\right)}_{m}\epsilon_{2,l}}}\kappa_{z}/\vert H_{2,y}\left(0\right)\vert\right).
\label{nl16}
\end{equation}
Note that the complete solution by including the plane wave term along \textit{x}-axis is $ H_{2,y}^{x,z} = H_{2,y}e^{ikx} $. 

The corresponding electric field components directed along the \textit{x} and \textit{z}-axes in the non-linear medium are
\begin{equation}
E_{2,x} = \dfrac{i\kappa_{z}}{\omega\epsilon_{0}\epsilon_{nl}}tanh\left[\kappa_{z}z + z^{'}\right]H_{2,y},
\label{nl6}
\end{equation}
and
\begin{equation}
E_{2,z} = -\dfrac{-k_{x}}{\omega\epsilon_{0}\epsilon_{nl}}H_{2,y}.
\label{nl7}
\end{equation}
\end{subequations}
For a numerical value of the non-linear dielectric constant of medium ``2" that appears in the electric and magnetic field equations, we set the average the magnetic field intensity to $ \vert\,H_{2,y}\left(0\right)\vert^{2}/2 $ to yield 
\begin{equation}
\epsilon_{nl} = \epsilon_{2,l} + \chi^{\left(3\right)}_{m}\vert\,H_{2,y}\left(0\right)\vert^{2}/2.
\label{nl8}
\end{equation}
The magnetic field, $ \textbf{H} $, at the interface can be easily determined for an incident radiation of given intensity; to carry out a numerical estimate, we suppose light of intensity, $ \mathcal{I} $, is incident on the sample. The incident light intensity can be equated to electromagnetic energy density~\cite{zangwill2013modern} as $ \mathcal{I} = \dfrac{\epsilon_{0}E^{2}c}{2} $ to give $ \vert H \vert = \sqrt{\dfrac{2\mathcal{I}}{\epsilon_{0}\mu_{0}^{2}c^{3}}} $, where $ c $ and $ \mu_{0} $ denote the speed of light and magnetic permeability in vacuum, respectively. The magnetic induction vector $ \textbf{B} = \mu_{0}\textbf{H} $ and $ E/B = c $. For the interface on which the light beam is incident, we rewrite the intensity relationship as $ \mathcal{I} = \dfrac{\epsilon_{avg}E^{2}c}{2} $, where $ \epsilon_{avg} = \dfrac{\epsilon_{1} + \epsilon_{2}}{2} $.

The reflection and transmission coefficients can now be easily obtained by resorting to Maxwell's boundary conditions. Again, enforcing the continuity of the tangential components at the interface and labelling the media on left (right) as ``1(2)", we get at $ z = 0 $
\begin{equation}
H_{1,y}^{i}\left(0\right) +  H_{1,y}^{r}\left(0\right) - H_{2,y}\left(0\right) = \sigma\,E_{2,x}\left(0\right), 
\label{hbc1}
\end{equation}
and
\begin{equation}
E_{1,x}^{i}\left(0\right) +  E_{1,x}^{r}\left(0\right) = E_{2,x}\left(0\right).
\label{ebc1}
\end{equation}
\label{tbc}
The superscripts ``i" and ``r" denote the incident and reflected components of electric and magnetic fields of the impinging light beam. Using the transfer matrix formalism, the amplitude of the magnetic fields on either side of the interface can be written as
\begin{subequations}
\begin{equation}
\begin{pmatrix} A_{1} \\
B_{1}
\end{pmatrix} =  M_{tm} \begin{pmatrix} A_{2} \\
B_{2}
\end{pmatrix}.
\label{tmat1}
\end{equation}
The transfer matrix $ M_{tm} $ in Eq.~\ref{tmat1} allows us to compute reflectance and transmittance amplitudes; in particular, reflectance and transmittance amplitudes are
\begin{eqnarray}
r = \dfrac{M_{tm}\left(2,1\right)}{M_{tm}\left(1,1\right)}, \label{ref} \\
t = \dfrac{1}{M_{tm}\left(1,1\right)}.
\label{trans}
\end{eqnarray}
\end{subequations}
The reflection and transmission coefficients are $ R = \vert r \vert^{2} $ and $ T = \vert t \vert^{2} $ which add to unity $\left(R + T = 1 \right)$ for zero absorption losses.
\noindent Carrying out the algebraic manipulation and using Eq.~\ref{trans}, the reflection (``r'') and transmission (``t") coefficients for a TM wave are
\begin{subequations}
\begin{equation}
r = \dfrac{k_{1z}/\epsilon_{1}\left[1 + \sigma\,\dfrac{i\kappa_{z}}{\omega\epsilon_{0}\epsilon_{nl}}tanh\left(\kappa_{z}z^{'}\right)\right] - \dfrac{i\kappa_{z}}{\epsilon_{nl}}tanh\left(\kappa_{z}z^{'}\right)}{k_{1z}/\epsilon_{1}\left[1 + \sigma\,\dfrac{i\kappa_{z}}{\omega\epsilon_{0}\epsilon_{nl}}tanh\left(\kappa_{z}z^{'}\right)\right] + \dfrac{i\kappa_{z}}{\epsilon_{nl}}tanh\left(\kappa_{z}z^{'}\right)} ,
\label{rcff}
\end{equation}
and
\begin{equation}
t = \dfrac{2k_{1z}/\epsilon_{1}}{k_{1z}/\epsilon_{1}\left[1 + \sigma\,\dfrac{i\kappa_{z}}{\omega\epsilon_{0}\epsilon_{nl}}tanh\left(\kappa_{z}z^{'}\right)\right] + \dfrac{i\kappa_{z}}{\epsilon_{nl}}tanh\left(\kappa_{z}z^{'}\right)}.
\label{tcff}
\end{equation}
\end{subequations}
Notice that the conductivity in Eqs.~\ref{rcff} and ~\ref{tcff} is a complex quantity of the form $ \sigma_{R} + i\sigma_{I} $, where $ \sigma_{R} $ and $ \sigma_{I} $ denote the real and imaginary components of the conductivity, respectively.

We quantitatively deduce the optical response by numerically evaluating the reflectance and trasmittance coefficients (Eqs.~\ref{rcff} and ~\ref{tcff})for the model structure in Fig.~\ref{schema}; additionally, we carry out a comparative study of the reflectance and transmittance coefficients with two different transition metal dichalcogenide sheets, in particular, MoS$_{2}$ and WSe$_{2}$ sheets. The choice of MoS$_{2}$ and WSe$_{2}$ mono-layer TMDCs lies in their widely-varying spin-orbit coupling splitting while the other dispersion parameters (Table.~\ref{table1}) are sufficiently close in magnitude. We explicitly discuss the role of the spin-orbit coupling induced spin split bands in mono-layer TMDCs in the following Section.~\ref{rd}.

\vspace{0.35cm}
\section{Numerical results}
\vspace{0.35cm}
\label{rd}
In the broad theoretical framework developed in the preceding sections, it is possible to perform a numerical assessment of the optical behaviour of a TMDC mono-layer inserted between a linear and non-linear medium/dielectric. To carry out calculations presented in this section, we first fix various parameters that define the arrangement; first of all note that the medium ``2'' on the right of the interface shown in Fig.~\ref{schema} is of the Kerr non-linear type whose third-order non-linear optical coefficient, following Ref.~\onlinecite{boyd2003nonlinear}, is of the order $ 10^{-8}\,m^{2}/V^{2} $. Further, the region to the left and right of the interface have dielectric constants of $ n_{1} = 3.0 $ and $ n_{2} = 2.0 + 0.05\textit{i} $, respectively and all media are taken to be non-magnetic $ \left(\mu = 1 \right) $. The Fermi level for the TMDC mono-layer is set to the top of the valence band which makes it effectively hole-doped (\textit{p}-type) and the angle of incidence is equal to $ 78\,^{\circ} $. We explicitly mention in the text if a different incident angle is chosen to aid the study of the described set-up. Note that the analysis assumes a TM-wave which imparts it a linear polarization. The case of a circularly-polarized incident beam will be qualitatively taken up later.

\begin{figure}
\begin{subfigure}
\centering
\includegraphics[scale=0.7]{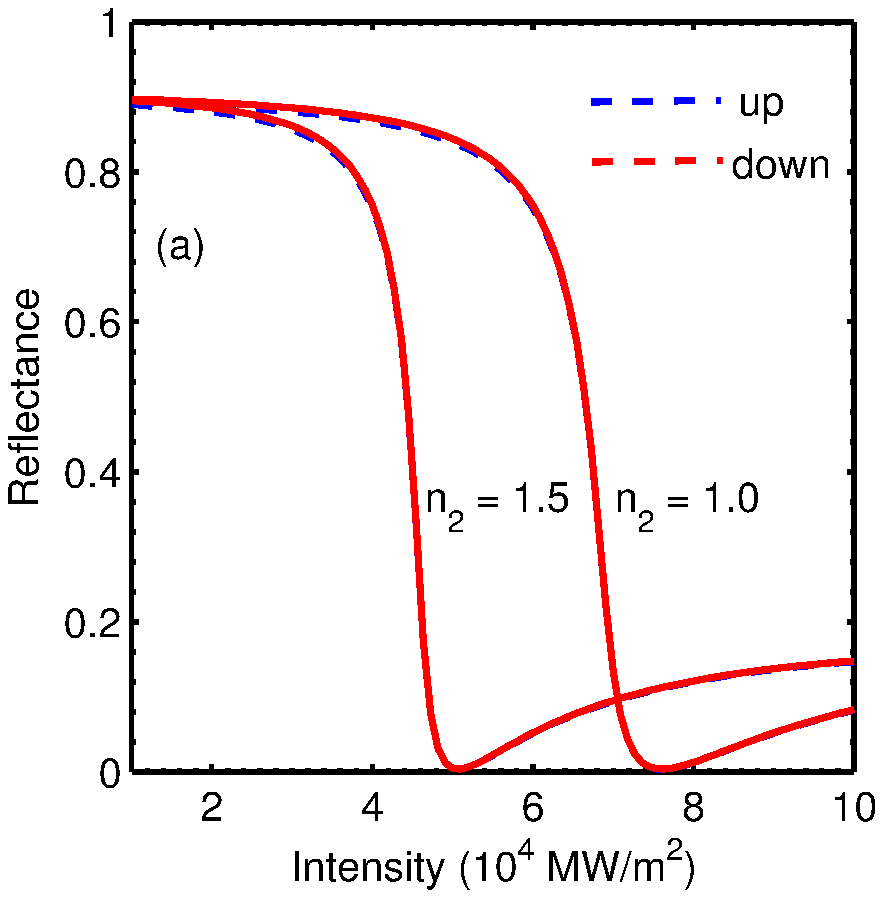} 
\end{subfigure}
\begin{subfigure}
\centering
\includegraphics[scale=0.7]{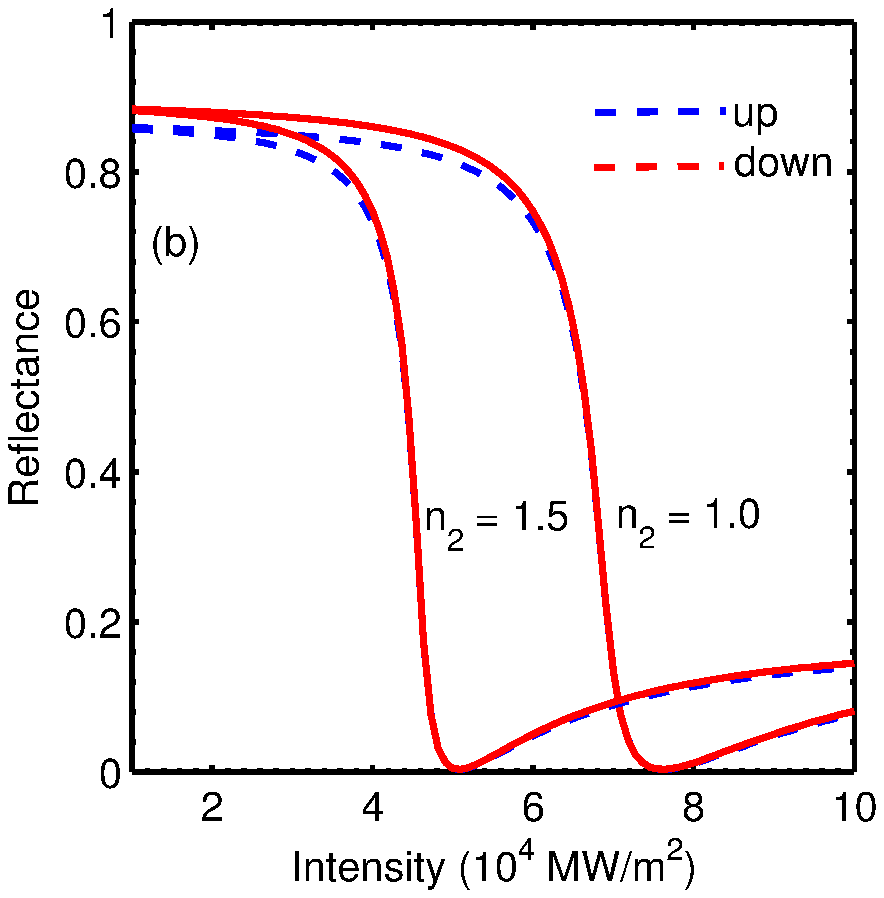}  
\end{subfigure}
\caption{The reflectance of mono-layer MoS$_{2}$(upper panel) and WSe$_{2}$ at $ T = 0\,K $. The reflectance close to  unity in both panels starts near total internal reflection and drops at the intensity-dependent critical angle before finite transmission begins. The two spin split valence bands for both MoS$_{2}$ and WSe$_{2}$ which give rise to distinct optical transition gaps (to the bottom of the conduction band do not exhibit a large difference in their reflectance coefficient. Note that the two different third-order non-linear susceptibility constants on the figure $ n_{2} $ are of the order $ 10^{-8}m^{2}/V^{2}$. They have been truncated for brevity. }
\label{reflpl}
\end{figure}

We also \textit{unequivocally} state that any reference to `spin' pertains to the spin degree of freedom of the electrons in the valence bands and has no direct bearing on the incident, reflected, and transmitted optical beams. To further elucidate the role of spin, the spin-orbit coupling splits the valence bands (see Fig.~\ref{eigvalpic}) while the bottom of the conduction band remains degenerate at $ K $; this furnishes two optical transition gaps  from top of the valence bands to bottom of the conduction band which we examine in this work. Furthermore, since we confine ourselves to light absorption in vicinity of the $ K $ or $ K^{'} $ edge of the Brillouin zone of the TMDC mono-layer, the optical transition gaps have been set to the following numbers.  For the MoS$_{2} $ (WSe$_{2}$) mono-layer, the optical transition gap between the top of the first valence band to bottom of the conduction band is $ 1.5850\, eV \left(1.37\,eV\right) $; correspondingly the energy difference for an electron making an inter-band transition from the top of the spin split lower valence band to bottom of the conduction band is set to $ 1.7350\, eV \left(1.83\,eV\right) $. We tacitly assume that an incident beam though of varying intensities carries an energy that is exactly equal to one of the transition gaps noted here. For the reflectance traces shown later, a reference to the `up (down)-spin' denotes that the incident beam was of energy that enabled an optical transition from top of the upper (lower) valence band to bottom of the conduction band. The inter-band optical conductivities (the real and imaginary parts) that correspond to these gaps along with the other parameters noted above can be simply plugged in Eq.~\ref{rcff} to furnish the desired reflectance (and transmittance) plots. The first set of reflectance plots shown in Fig.~\ref{reflpl} contain several note-worthy features that we discuss below.

Firstly, the reflection and transmittance when light is incident from a denser to rarer medium are a function of the intensity of the incident radiation which is a direct outcome of the refractive index of the second optical medium has an electric/magnetic field (and consequently intensity-dependent) controlled non-linear component. An additional direct manifestation of this intensity dependence is also seen in the critical angle; for the simple case of reflection and transmission between two distinct optical media (the refractive indices $ are n^{1} $ and $ n^{2} $) with linear dielectrics, the critical angle is $ \theta_{cr} = sin^{-1}\left(n^{1}/n^{2}\right) $, which, however, isn't true for our case. Finally, we note that the reflectance coefficients in Fig.~\ref{reflpl} for the selected parameters begin close to unity, display a sharp drop and rise again. Further, for two set of values of the non-linear coefficient, $ n_{2} $, the drop in reflectance happens at different intensity levels. The initial value of the reflectance starts from close to unity (mirroring total internal reflection) at low intensities since the chosen angle of incidence ($ \theta_{i} = 78\,^{\circ} $) exceeds the critical angle; the critical angle increases with higher intensities but continues to lie below $ \theta_{i} $ until at a threshold intensity, the equality $ \theta_{i} =  \theta_{cr} $ is satisfied. The threshold intensity marks the end of total internal reflection, the reflectance abruptly drops to a low value beyond which it rises again as the arrangement of Fig.~\ref{schema} begins to partially reflect and transmit light. The non-linear coefficient $ n_{2} $ for two different values as indicated in Fig.~\ref{reflpl} gives a different critical angle in each case which is borne out by the dip in the reflectance coefficient at two discrete intensities. As a possible application of this behaviour, we can conceive of applications where it is expedient to harness the intensity of light as an external tunable parameter and modulate the flow of light.

We also like to comment on the role of spin-orbit coupling in splitting the valence bands that yield two optical band gaps as noted at the beginning of this section. Note that the optical band gap is one of the material parameters that regulates the inter-band optical conductivity, which based on Eqs.~\ref{rcff} and ~\ref{tcff}, determines the final reflectance and transmission coefficients. The reflectance traces for both the TMDC mono-layers corresponding to the two optical gaps (marked as \textit{up} and \textit{down} on Fig.~\ref{reflpl}) are indistinguishable for over the full range of incident intensities which allows us to conclude that the complex inter-band conductivities are sufficiently close. Further, the relatively large spin-splitting in WSe$_{2}$ (three-fold higher than MoS$_{2}$) notwithstanding, the reflectance coefficients when a mono-layer of this TMDC is inserted at the interface of linear and non-linear dielectric is very comparable to those obtained with MoS$_{2}$. The higher spin-orbit coupling in WSe$_{2}$ does not translate in to a significant increase in the inter-band optical conductivity. In fact, to test our reasoning, we artificially augmented the conductivity of WSe$_{2}$ by a factor of three-halves to observe a clear increase in the attendant reflectance plots over their  MoS$_{2}$ counterparts. It therefore suffices to say that influences which reinforce or impede the inter-band conductivity serve to moderate the optical response of the TMDC mono-layer. The spin-orbit coupling for semiconducting TMDCs is, however, not such an influence.

\begin{figure}
\includegraphics[scale=0.8]{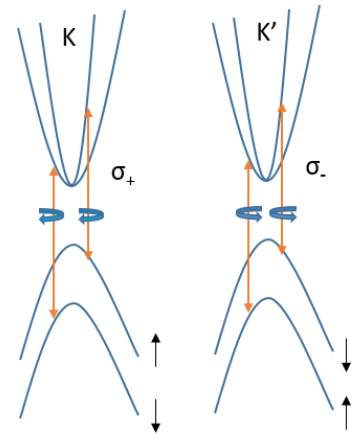}
\caption{A pictorial description of the valley edges $ K $ and $ K^{'} $ in the Brillouin zone of a mono-layer TMDC with opposite spins. The valley edges undergo a selective optical transition when an incident beam of definite circular-polarization shines on the mono-layer. A right- (left-) circularly polarized beam excites an electron from the $ K (K^{'}) $ edge. Note that this spin and valley optical selection rule is governed by the principle of conservation of angular momentum and time reversal symmetry.}
\label{tvps}
\end{figure}

An additional note about the polarization of incident light is in order here: The $ K $ and $ K^{'} $ edge of the Brillouin zone in mono-layer TMDC are energy degenerate and time-reversed counterparts. A circularly polarized light aligns with a specific spin (see Fig.~\ref{tvps}) to excite optical transition exclusively at one of the valley edges, for instance, a right circularly-polarized light will excite electrons at $ K $ but not $ K^{'} $. Since we have chosen a linearly-polarized beam which in principle can be decomposed in a right- and left-circularly polarized component, optical transitions happen at both $ K $ and $ K^{'} $. A liearly-polarized beam of frequency that matches one of the transition gaps will promote an electron to the conduction state at both valley edges.

\begin{figure}
\includegraphics[scale=0.7]{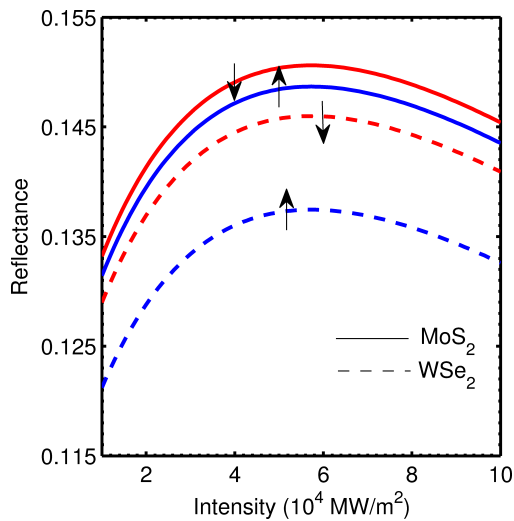}
\caption{The reflectance coefficients for a series of intensity levels when light is incident from a rarer to the denser Kerr non-linear medium. The angle of incidence is, as before, set to $ \theta_{i} = 78\,^{\circ} $. The arrows on the plot refer to spin-up and spin-down and point to the reflection coefficients for an optical transition that may happen from the top of the lower (upper) spin-down (up) valence band to bottom of the conduction band at $ K $ or $ K^{'} $ valley-edge of the Brillouin zone.}
\label{rtdpl}
\end{figure}

The preceding analysis of intensity-modulated reflection and transmission can also be carried out when the incident light beam travels from a rarer to denser medium. For the arrangement shown in Fig.~\ref{schema}, this turns the non-linear dielectric in to a the denser medium. The phenomenon of total internal reflection is absent and the reflectance plots therefore do not display a drop (Fig.~\ref{rtdpl}) but instead show a declining trend with increased intensity levels.  Note that a higher intensity level enhances the non-linear refractive index and the medium is more optically dense. For a numerical answer, we set $ n_{1} $ to 1.5 while  $ n_{2} = 2.0 + 0.05\textit{i} $ remains unaltered.

For a targeted application, the optical response can be externally tuned either by a careful impregnation of impurities at the time of growth or suitable structural modifications. As for the broad structural and electronic behaviour of mono-layer TMDC, they are generally well reflected by the set of parameters given in Table.~\ref{table1}. While the spin-orbit coupling is intrinsic to the material, determined primarily by size of the nucleus and thus immune to most perturbations including the often employed external strain, the other band parameters listed in Table.~\ref{table1} could be suitably amended. In fact, strain-engineering has been carried out previously on TMDC devices~\cite{tabatabaei2013first,hosseini2015strain} and predict improved performance characteristics. In our work, we utilize results from a first-principles based calculation by Mandiaki \textit{et al.} who explicitly tabulate the change in energy band gap (and the dielectric constants) for mono-layer TMDCs.~\cite{maniadaki2015strain} One of their findings shows that mono-layer MoS$_{2}$ transitions to an indirect semiconductor for isotropic and uniaxial strain while WSe$_{2}$ retains its direct band gap character under tensile isotropic and low values of uniaxial strain. As an illustration of the possibility to engineer optical response through external strain, we consider an isotropic and $ 3\% $ tensile-strained WSe$_{2}$ sheet. The relevant direct band gap at $ K $ for the strained case is $ 1.53\,eV $ (the pristine value is $ 1.60\,eV $) and the hopping parameter $ t $ in Eq.~\ref{mos2ham} is empirically scaled down by the factor $ 1.53/1.60 $ to $ 1.137\,eV $ while the new lattice constant is  $ 4.303\,\AA $. 

\begin{figure}
\includegraphics[scale=0.75]{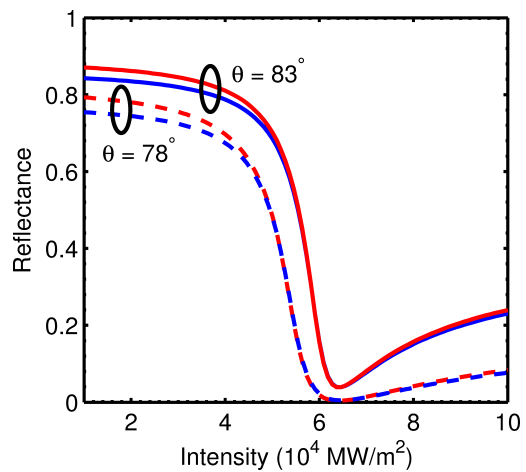}
\caption{The reflectance coefficients with strained WSe$_{2} $ is displayed for two angles of incidence, $ \theta_{i} = 78\,^{\circ} $ and $ \theta_{i} = 83\,^{\circ} $.  The introduction of strained WSe$_{2}$ readjusts the critical angle as compared to the unstrained case shown in the bottom panel of Fig.~\ref{reflpl} at which the system switches from total internal reflection to normal transmission. The optical parameters for this figure are the ones defined at the beginning of Section.~\ref{rd}.} 
\label{reflstr}
\end{figure}

We plot the reflectance coefficient in Fig.~\ref{reflstr} for this case and immediately notice that it is lower than the corresponding unstrained mono-layer WSe$_{2} $ and the transition to normal transmission from total internal reflection happens at a higher  incident intensity. The reflectance is restored when the angle of incidence is changed to $ 83^{\circ} $ though the threshold intensity at which total internal reflection ends is higher than the unstrained case. We have noted before that the intensity-governed Kerr non-linearity determines the critical angle at which normal transmission takes over from total internal reflection; however, the presence of a TMDC sheet by virtue of its surface conductivity which changes under strain influences this shift as evident from Fig.~\ref{reflstr}. The surface conductivity, therefore, in addition, to incident light intensity, and the Kerr coefficient of non-linearity allows an external adjustment of the passage of light through the proposed arrangement of Fig.~\ref{schema}.

\vspace{0.35cm}
\section{Summary}
\vspace{0.35cm}
We have theoretically obtained the behaviour of a linearly-polarized light beam incident at the interface of a linear and non-linear dielectric. In addition, the interface is covered by a sheet of mono-layer TMDC that mediates the transmission of light through its inter-band optical conductivity. The final reflectance and transmittance coefficients were determined by resorting to a transfer matrix approach and the inter-band optical conductivity of mono-layer TMDC was quantitatively calculated with the Kubo formalism. Significantly, for a beam of light traveling from an optically denser medium with a linear dielectric behaviour to an optically rarer Kerr-type region, a threshold intensity level exists for which the system switches from total internal reflection to normal transmission. This switch is a reminder that for the Kerr-type dielectric described by Eq.~\ref{knl}, the critical angle is not a simple reproduction of Snell's law but instead is intensity-dependent. No such switch is observed though for a light beam that traverses an optically rarer medium with linear dielectric in to the denser Kerr-region. Further our choice of two different mono-layers of TMDCS, MoS$_{2}$ and WSe$_{2}$, because of their near-identical band parameters but widely varying spin-orbit coupling does not display any noticeable change in the reflectance pattern. The spin-orbit coupling in WSe$_{2}$ though three-fold higher than MoS$_{2}$ is a relatively low number to significantly impact the inter-band conductivity and the overall reflectance pattern. However, we find that external perturbations such as mechanical strain in mono-layer TMDCs with adjustments to the angle of incidence can tune the light propagation in our setup, reflected in a shift of the reflectance coefficients. Further, other variants of strain as dominant perturbations including uniaxial and shear in addition to the tensile strain utilized in this work, in principle, allow us to adapt the optical response through an alteration of their inter-band optical conductivity.

The use of a non-linear medium also allows us to examine several optical phenomena that we haven't explored in this work. In particular, the phenomenon of optical bistability~\cite{dai2015tunable}  by which an optical medium has two possible optical outputs for a unique optical input, holds promise for the efficient integration of high-speed photonic circuitry to nanoscale electronics and a full optical control over switches~\cite{tamm1990bistability}, logic gates, and memory chips.~\cite{firth1988optical} A simple way to observe non-linearity is through introduction of a functional non-linear material with an intensity-dependent refractive index. This phenomenon to which we have alluded in Section.~\ref{kerrd} is the Kerr effect and is an inherently weak effect requiring considerable power densities to fully manifest. In this work, while we haven't attributed any non-linearity to the TMDC mono-layer that covers the interface of a linear and Kerr- (non-linear) type dielectric and chosen to focus on the mediation of the passage of light by virtue of its inter-band surface conductivity, the 2D surface of TMDC, however, can significantly localize light~\cite{huang2015optical} and enhance the optical fields. We believe it would be a highly compelling study to investigate the non-linearity of TMDCs inside optical resonating cavities of the Fabry-Perot type. Before closing, it must be mentioned that while we have used the semiconducting mono-layer MoS$_{2}$ and WSe$_{2}$ as typical representatives of two-dimensional layered transition metal dichalcogenides from a wide range of available compounds,~\cite{friend1987electronic, wilson1969transition} it would certainly be worthwhile to assess the optical response of other recently discovered two-dimensional materials such as phosphorene and stanene.

This work was supported in part by the BU Photonics Center and U.S. Army Research Laboratory through the Collaborative Research Alliance(CRA).

\bibliographystyle{apsrev}
%\bibliography{References}

\begin{appendices}
\appendix
\renewcommand \thesubsection{\Roman{subsection}}
\titlespacing\section{5pt}{12pt plus 4pt minus 2pt}{0pt plus 2pt minus 2pt}

\vspace{0.3cm}
\section{Wave propagation in non-linear Kerr media}
\vspace{0.3cm}
We derive the weakly non-linear Helmholtz equation (cf. Eq.~\ref{nl2}) for a propagating light wave in a medium with Kerr non-linearity. For low electric fields \textbf{E}, to a first order approximation, the polarization field can be represented through a linear equation
\begin{subequations}
\begin{equation}
\textbf{P}_{lin} = \epsilon_{0}\chi^{\left(1\right)}_{e}\left(\omega\right)\textbf{E},
\label{pf1} 
\end{equation}
where $ \chi_{e}^{\left(1\right)}\left(\omega\right) $ is the frequency-dependent first-order susceptibility tensor. For ease of notation, we simply denote the electric susceptibility tensor of any rank, $ \chi_{e}^{\left(m\right)}\left(\omega\right)$, as $ \chi_{e}^{\left(m\right)} $ though an implicit frequency-dependence is tacitly assumed. With this in mind, the electric displacement vector, \textbf{D} is 
\begin{equation}
\textbf{D} = \epsilon_{0}\textbf{E} + \textbf{P}_{lin} = \epsilon_{0}\left(1 + \chi^{\left(1\right)}_{e}\right)\textbf{E}.
\label{pf2} 
\end{equation}
Note that $ n_{0}^{2} = 1 + \chi $ is the linear refractive index of the medium which allows us to recast the Gauss electrostatic equation as
\begin{equation}
\textbf{D} = \epsilon_{0}n_{0}^{2}\left(\omega\right)\textbf{E}.  
\end{equation}

\noindent The Helmholtz equation is derived from Maxwell's equation to give 
\begin{flalign}
\nabla^{2}\textbf{E} -\dfrac{1}{c^{2}}\dfrac{\partial^{2}}{\partial t^{2}}\textbf{E} = 0, \notag \\
\nabla^{2}\textbf{E} + \dfrac{\omega^{2}}{c^{2}}\textbf{E} = 0.
\label{pf3}
\end{flalign}
The operator $ \dfrac{\partial^{2}}{\partial\,t^{2}} $ is replaced by $ -\omega^{2} $ for a sinusoidal wave. Under conditions where the electric displacement vector \textbf{D} is modified as shown in Eq.~\ref{pf2}, the Helmholtz equation, starting from Maxwell's equation takes the form
\begin{equation}
\nabla^{2}\textbf{E} + \dfrac{\omega^{2}n_{0}^{2}}{c^{2}}\textbf{E} = 0.
\label{pf4}
\end{equation}
Eq.~\ref{pf4} differs from Eq.~\ref{pf3} by an additional factor of $ n_{0}^{2} $ that precedes the \textbf{E} vector.

For the case of an isotropic material with weak Kerr non-linearity, Eq.~\ref{pf1} has a non-linear component $ \textbf{P}_{nl} $, which under a Taylor expansion will have additional higher order odd terms. All even ordered susceptibility tensor terms on account of symmetry must vanish~\cite{dragoman2013optical}, leaving us with an expression of the form
\begin{equation}
\textbf{P}_{nl} = \chi^{\left(3\right)}_{e}\left(\omega\right)\textbf{E}^{3} + \chi^{\left(5\right)}_{e}\left(\omega\right)\textbf{E}^{5} + ...
\label{pf5}
\end{equation}
Retaining only the cubic term, the non linear polarization, for a sinusoidal electric field $ \textbf{E} = E\,exp(-i\omega\,t)+ c.c $ is now
\begin{equation}
\textbf{P}_{nl} = \chi^{\left(3\right)}_{e}\left(3\vert E \vert^{2}E\,exp\left(-i\omega\,t\right) + E^{3}exp\left(-3i\,\omega\,t\right) + c.c \right). 
\label{pf6}
\end{equation}
Making a further approximation, we ignore the third harmonic component since it is phase mismatched with the first harmonic component; Eq.~\ref{pf6} is therefore
\begin{flalign}
\textbf{P}_{nl} &= \chi^{\left(3\right)}_{e}\left(3\vert E \vert^{2}E\,exp\left(-i\omega\,t\right) + c.c \right), \notag \\
&= 3\chi^{\left(3\right)}\vert E \vert^{2}\textbf{E}.
\label{pf7} 
\end{flalign} 
\end{subequations}
The non linear term in Eq.~\ref{pf7} transforms the refractive index $ n_{0}^{2} = 1 + \chi_{e}^{\left(1\right)} $ to 
\begin{equation}
n^{2} = n_{0}^{2} + \alpha\,\vert E \vert^{2}, 
\label{nonkerr}
\end{equation}
which when inserted in Helmholtz equation (Eq.~\ref{pf4}) gives $ \nabla^{2}\textbf{E} + \dfrac{\omega^{2}n^{2}}{c^{2}}\,\textbf{E} = 0 $. The constant $ \alpha $, when only the third rank susceptibility tensor is included can be expressed as $ \xi\,\chi_{e}^{\left(3\right)} $, where $ \xi $ is an experimentally determined constant. Under conditions of weak non-linearity, Eq.~\ref{nonkerr} can be further simplified to
\begin{equation}
n = n_{0} + n_{2}\vert\,E\vert^{2},
\end{equation}
which is what we originally started with in Eq.~\ref{knl}.

The analogous equation for the $ \textbf{H} $ field induced non-linearity can be derived by considering the third order magnetic susceptibility tensor $ \chi_{m}^{\left(3\right)}\left(\omega\right)$. For our case, the magnetic field vector along \textit{y}-axis and propagating along the \textit{z}-axis, at a certain frequency acts on itself through the non linearity term~\cite{boyd2003nonlinear} $ \epsilon\,\chi_{m}^{\left(3\right)}\vert H_{y} \vert^{2}\,H_{y} $ to give 
\begin{equation}
\dfrac{\partial^{2}H_{y}}{\partial\,z^{2}} + \dfrac{\omega^{2}}{c^{2}}\left(\epsilon + \epsilon\,\chi_{m}^{\left(3\right)}\vert H_{y} \vert^{2} - k_{x}^{2}\right)H_{y} = 0.
\label{pf8}
\end{equation}
Eq.~\ref{pf8} is reduced to Eq.~\ref{nl2} under the substitution $ k_{x}^{2} = \left(\omega/c\right)^{2}\epsilon\,sin^{2}\theta_{1} $. Note that $ \epsilon $ is the dielectric constant(the linear part) of the non-linear medium, $ \theta_{1} $ is the angle of incidence, and the usual dispersion relation $ k_{x}^{2} + k_{x}^{2} = \dfrac{\omega^{2}}{c^{2}}n^{2} $ holds in this medium.

\vspace{0.3cm}
\section{Conductivity of mono-layer transition metal dichalcogenides}
\vspace{0.3cm}
We include steps left out in the main text to arrive at the $ T = 0 \, K $ conductivity expression shown in Eq.~\ref{rsigxx1}. The conductivity expression using Kubo formula which we write here again is
\begin{equation}
\sigma_{x,y} = -i\dfrac{\hbar\,e^{2}}{L^{2}}\sum_{n,n^{'}}\dfrac{f\left(\varepsilon_{n}\right)- f\left(\varepsilon_{n^{'}}\right)}{\varepsilon_{n} - \varepsilon_{n^{'}}}\dfrac{\langle\,n\vert\,v_{\alpha}\vert\,n^{'}\rangle \langle\,n^{'}\vert\,v_{\alpha}\vert\,n\rangle}{\varepsilon_{n} - \varepsilon_{n^{'}}+i\,\zeta},
\label{kubofa}
\end{equation}
The matrix element, $ M_{1} = \langle\,n\vert\,\hat{v}_{\alpha}\vert\,n^{'}\rangle $, that appears in Eq.~\ref{kubofa} for $ \sigma_{xx} $ becomes
\begin{eqnarray}
M_{1} &=& \dfrac{at}{2\hbar}\begin{pmatrix}
\eta_{+}e^{i\theta} & \eta_{-} 
\end{pmatrix}\begin{pmatrix}
0 & 1 \\
1 & 0
\end{pmatrix}\begin{pmatrix}
\eta_{-}e^{-i\theta} \\
-\eta_{+}
\end{pmatrix}, \notag \\
&=& \dfrac{at}{2\hbar}\left[-\eta^{2}_{+}e^{i\theta} + \eta^{2}_{-}e^{-i\theta}\right], 
\label{matdev}
\end{eqnarray}
where 
\begin{equation}
\eta_{\pm}^{2} = 1 \pm \dfrac{\Delta \pm \lambda}{\sqrt{\left(\Delta \pm \lambda\right)^{2} + \left(2atk\right)^{2}}}.
\label{etaapp}
\end{equation}
The velocity operator along the \textit{x}-axis is $ \hat{v}_{x} = \left(at/\hbar\right)\hat{\sigma}_{x}$. Also, $ \Delta - \lambda $ corresponds to optical band gap between spin-up states while for the case of spin-down bands, we simply need to make the change $ \Delta - \lambda \rightarrow \Delta + \lambda $. The wave functions given in Eq.~\ref{wfs1} have been used in Eq.~\ref{matdev}. Evaluating the expression for the matrix product in Eq.~\ref{matdev} gives
\begin{equation}
M_{1} = -at\left(\Upsilon\,cos\theta + isin\theta\right), 
\label{matdevr}
\end{equation}
where 
\begin{equation}
\Upsilon = {\dfrac{\Delta \pm \lambda}{\sqrt{\left(\Delta \pm \lambda\right)^{2} + \left(2atk\right)^{2}}}}.
\label{constu}
\end{equation}
Further, since the other matrix element, $ M_{2} = \langle\,n^{'}\vert\,\hat{v}_{\alpha}\vert\,n\rangle $ is just $ M_{1}^{*} $, the product $ M_{1} \times M_{2} $ in the Kubo expression(Eq.~\ref{kubofa1}) becomes
\begin{equation}
M_{1} \times M_{2} = a^{2}t^{2}\left(\Upsilon^{2}cos^{2}\theta + sin^{2}\theta\right),
\label{matprod}  
\end{equation}
As stated before, the upper(lower) sign stands for optical transition between the spin-up(down) states. Substituting the matrix products in Eq.~\ref{kubofa} and expanding the sum, one obtains 
\begin{widetext}
\begin{equation}
\sigma_{xx} = -i\dfrac{\hbar\,e^{2}}{4\pi^{2}}\int_{0}^{k_{c}}\,dk\,\int_{0}^{2\pi}\,d\theta\,\dfrac{f\left(\varepsilon_{c}\right)- f\left(\varepsilon_{v}\right)}{\varepsilon_{c} - \varepsilon_{v}}a^{2}t^{2}\left(\Upsilon^{2}cos^{2}\theta + sin^{2}\theta\right)\biggl[\dfrac{1}{\hbar\,\omega + \Delta\,\varepsilon + i\zeta} 
+ \dfrac{1}{\hbar\,\omega - \Delta\,\varepsilon + i\zeta}\biggr].
\label{kubofa1}
\end{equation}
\end{widetext}
The summation has been replaced by an integral in 2D $ k $-space using $ \sum \rightarrow \dfrac{\mathcal{A}}{4\pi^{2}}\int\,d^{2}k $, where $ \mathcal{A} $ is the sample area. Note that the optical transitions happen between the valence and conduction band state, we have therefore replaced the variables $ \varepsilon_{n} $ and $ \varepsilon_{n^{'}} $ by $ \varepsilon_{c} $ and $ \varepsilon_{v} $, respectively in Eq.~\ref{kubofa1}. 

\noindent For a clean non-interacting sample, $ \zeta \rightarrow 0 $, using the identity $ \dfrac{1}{x + i0^{+}} = \mathbb{P}\dfrac{1}{x} - i\pi\delta\left(x\right)$, lets us separate the integral in Eq.~\ref{kubofa1} to write the real and imaginary components of the longitudinal conductivity. Since we perform calculations at $ T = 0\, K $, the derivation assigns the Fermi distribution to behave as a Heaviside function, $ \Theta\left(\cdot\right) $, such that $ f\left(\varepsilon_{c}\right) = 0 $ and  $ f\left(\varepsilon_{v}\right) = 1 $. Lastly, we substitute the energy band gap, $ \Delta\,\varepsilon $, from Eq.~\ref{dele} in Eq.~\ref{kubofa1} to obtain the final form of the longitudinal conductivity(Eq.~\ref{rsigxx1}).   

\end{appendices}

\end{document}